\documentclass[a4paper,12pt]{article}
\usepackage{graphics}
\usepackage{amsmath,amssymb}
\def\bk{{\bf k}}
\def\bp{{\bf p}}
\def\vf{\varphi}

\def\th{\theta}
\def\half{\frac{1}{2}}

\def\be{\begin{equation}}
\def\ee{\end{equation}}
\def\bea{\begin{eqnarray}}
\def\eea{\end{eqnarray}}
\def\beax{\begin{eqnarray*}}
\def\eeax{\end{eqnarray*}}
\title
{Equivalent Hamiltonian for Lee Model}
\author{H.~F.~Jones\thanks{h.f.jones@imperial.ac.uk}\\
Imperial College London}
\date{}

\begin{document}
\maketitle
\begin{abstract}
Using the techniques of quasi-Hermitian quantum mechanics and quantum field theory
we use a similarity transformation to construct an equivalent Hermitian Hamiltonian for the Lee model.
In the field theory confined to the $V/N\theta$ sector it effectively decouples $V$, replacing the three-point interaction of the original Lee model by an additional mass term for the $V$ particle and a four-point
interaction between $N$ and $\theta$. While the construction is originally motivated
by the regime where the bare coupling becomes imaginary, leading to a ghost, it applies equally
to the standard Hermitian regime where the bare coupling is real. In that case the similarity
transformation becomes a unitary transformation.
\\


\end{abstract}
\section{Introduction}
Following the original paper by Bender and Boettcher\cite{BB} on $PT$-symmetric but non-Hermitian
Hamiltonians, subsequent research initially concentrated on an exploration of the reality
or otherwise of the spectrum of non-Hermitian generalizations of well-known soluble models (see
Ref.~\cite{GZ} for a systematic approach).
However, it soon became apparent that something additional to the reality of the spectrum was
needed if such models were to be viable as quantum theories with a proper probabilistic interpretation.
This is because the natural metric for $PT$-symmetric models gives the overlap of two wave-functions
$\psi(x)$ and $\vf(x)$ as $\int dx \vf^*(-x) \psi(x)$, rather than the usual $\int dx \vf^*(x) \psi(x)$.
Since the corresponding norm is not positive definite, the theory endowed with this metric does not
represent a physical framework for quantum mechanics. Instead it turns out to be possible to
construct\cite{BBJ, Hendrik, AM-metric}
an alternative, positive-definite metric $\eta\equiv e^{-Q}$, which is dynamically determined by the particular
Hamiltonian in question, satisfying
\bea\label{qH}
H^\dag=e^{-Q} H e^{Q}.
\eea
$H$ is then said to be quasi-Hermitian with respect to $\eta$.
It was also shown \cite{AM-h} that $\rho=\sqrt{\eta}$ provided a
similarity transformation from the non-Hermitian $H$ to an equivalent Hermitian $\tilde{H}$, namely
\bea\label{h}
\tilde{H}=e^{-Q/2} H e^{Q/2},
\eea
and this equivalent
Hermitian Hamiltonian was subsequently constructed, frequently in perturbation theory only, in a variety
of models\cite{HFJ-ix3, AM-ix3, -x4}.

These techniques were used by Bender et al.\cite{Ghost} to show that the ghost state that appears in the
Lee model\cite{Lee, Schweber, Barton}
when the renormalized coupling constant exceeds a critical value can be treated as a viable state
with a positive norm if the appropriate metric is used. Since they calculated this metric in the quantum mechanical version
and in the $V-N\theta$ sector of the field theory, a natural extension of their work is to
attempt to calculate the equivalent Hermitian Hamiltonians in each case, obtained by the similarity transformation
alluded to above. That is the task of the present paper.

In Section 2 we present the quantum mechanical calculation, which gives a closed-form expression for the
equivalent Hermitian Hamiltonian. The field theory problem is tackled in Section 3, confined to the $V-N\theta$
sector. We discuss the significance of these results in Section 4, noting in particular that the construction
applies equally to the situation where the renormalized coupling is less than the critical value and the original
Hamiltonian is also Hermitian.

\section{Quantum Mechanics}
\setcounter{footnote}{1}
\renewcommand{\thefootnote}{{\fnsymbol{footnote}}}
The Hamiltonian for the simplified quantum-mechanical version of the Lee model is
\bea
H=H_0+H_1,
\eea
where
\bea\label{H1QM}
H_0&=&m_{V_0}V^\dag V +m_N N^\dag N +m_\th a^\dag a,\cr
&&\\
H_1&=&g_0(V^\dag N a +a^\dag N^\dag V).\nonumber
\eea
Here $N$ and $V$ are treated as fermions, with occupation number 1 or 0.
The bare states in the $V/N\th$ sector are denoted by $|1,0,0\rangle$
and $|0,1,1\rangle$, where the indices are the occupation numbers of
the bare $V$, $N$ and $\th$ particles respectively.

The energy eigenstates in the sector are given as linear combinations
\bea
|E\rangle=\alpha_1 |1,0,0\rangle +\alpha_2 |0,1,1\rangle,
\eea
with the coefficients $\alpha_i$ given by the solutions of the secular
equations
\bea
\left(\begin{array}{cc} m_{V_0}-E&g_0 \\ g_0 & m_N+m_\th -E\end{array}\right)\left(\begin{array}{c} \alpha_1\\ \alpha_2\end{array}\right)=0
\eea
Introducing the variable
\bea\label{x}
x\equiv \tan^{-1}\left(\frac{2g_0}{\mu_0}\right),
\eea
the two solutions can be written, for real $g_0$, as
\bea\label{evs}\begin{array}{llll}E\equiv m_V = m_{V_0}-g_0\tan\half x &
\mbox{with} & \alpha_1=\cos\half x, & \alpha_2=-\sin\half x\\
\mbox{and}&&&\\
E\equiv E_{N\th} = m_N+m_\th +g_0\tan\half x &
\mbox{with} & \alpha_1=\sin\half x, & \alpha_2=\cos\half x\end{array}
\eea
Explicitly
\bea\label{sqrt}
g_0\tan\half x = \half \left( \sqrt{\mu_0^2+4g_0^2}-\mu_0\right).
\eea
In this notation the renormalized coupling constant is given by $g=\mu\sin\half x$, where
$\mu\equiv m_N+m_\th-m_V$, and the relation between the bare and renormalized coupling
constants by $g=g_0\cos\half x$, so that
\bea
g_0^2=\frac{g^2}{1-g^2/\mu^2}.
\eea
A similar equation appears in the field theory version, where a ghost appears when $g$ exceeds a critical
value and
$g_0$ becomes pure imaginary. However, contrary to what is stated in Ref.~\cite{Ghost}, the situation is rather
different in quantum mechanics. Here the regime $g>\mu$ is not viable, because it produces a complex spectrum.

Thus, as $g$ passes through $\mu$, the unrenormalized coupling constant $g_0$ goes to infinity and then becomes imaginary,
with $|g_0|$ very large. In that regime, assuming that $\mu_0$ is real, it is clear from Eqs.~(\ref{evs}) and (\ref{sqrt})
that $m_V$ (and hence $\mu$) is complex, contrary to supposition. The only loophole, $\mu_0$ becoming complex, is ruled
out by the equation
\bea
\mu_0=\mu-\frac{g_0^2}{\mu},
\eea
which can easily be derived from the above equations.

Nonetheless, we can proceed with the basic framework of Ref.~\cite{Ghost}, but staying within
the allowed (Hermitian) regime $g<\mu$, so that $g_0$ and $x$ are both real. In that paper
the equation for the operator $Q$, namely
\bea\label{qH1}
[e^Q, H_0] = \{e^Q, H_1\},
\eea
which is equivalent to the condition (\ref{qH}) for $g_0$ imaginary,
is solved, with the result that
\bea\label{QQM}
Q=V^\dag N a \frac{1}{\sqrt{n_\th}}\ x_{n_\th}-\frac{1}{\sqrt{n_\th}}\ x_{n_\th}a^\dag N^\dag V,
\eea
where, in analogy with Eq.~(\ref{x}) we have introduced the operator
\bea\label{xth}
x_{n_\th}\equiv\tan^{-1}\left(\frac{2g_0 n_\th}{\mu_0}\right).
\eea
Here $n_\th$ is the number operator for the $\th$ particle: $n_\th=a^\dag a$.

Because the $V$ and $N$ particles are treated as fermions, with the simplification
that $n^2_{V,N}=n_{V,N}$ where $n_V$ and $n_N$ are the respective number operators,
the series for $e^Q$ can be summed exactly, with the result that, in our
notation\footnote{We have corrected a couple of errors in Eq.~(32) of \cite{Ghost}}
\bea
e^Q&=&1-n_N(1-n_V)(1-\cos x_{n_\th})-n_V(1-n_N)(1-\cos x_{n_\th+1})\cr&&\cr
&&+V^\dag N a \frac{1}{\sqrt{n_\th}} \sin x_{n_\th} - \sin x_{n_\th} \frac{1}{\sqrt{n_\th}}a^\dag N^\dag V,
\eea
which can be usefully rewritten as
\bea
e^Q=1-f_1 n_N\bar{n}_V-f_2 n_V\bar{n}_N +V^\dag N a f_3 -f_3 a^\dag N^\dag V,
\eea
where $\bar{n_{V,N}}\equiv 1-n_{V, N}$ and
\bea
f_1&=&1-\cos x_{n_\th},\nonumber\\
f_2&=&1-\cos x_{n_{\th+1}},\\
f_3&=&\frac{1}{\sqrt{n_\th}}\sin x_{n_\th} .\nonumber
\eea
We have independently verified this result using the condition
\bea\label{grev}
H(-g_0)=e^{-Q}H(g_0) e^Q
\eea rather than (\ref{qH1}).
In so doing, the identities
\bea\label{id}
f_2&=&f_1(n_\th+1),\nonumber\\
n_\th f_3^2&=&2f_1-f_1^2,\\
f_3&=& (2g_0/\mu_0) \cos x_{n_\th}\nonumber
\eea
proved crucial.

To construct the equivalent Hermitian Hamiltonian $\tilde{H}$ we need $e^{\half Q}$, which
can be expressed as
\bea
e^{\half Q}=1-\tilde{f}_1 n_N\bar{n}_V-\tilde{f}_2 n_V\bar{n}_N
+V^\dag N a \tilde{f}_3 -\tilde{f}_3 a^\dag N^\dag V,
\eea
where
\bea
\tilde{f}_1&=&1-\cos \half x_{n_\th},\nonumber\\
\tilde{f}_2&=&1-\cos \half x_{n_{\th+1}},\\
\tilde{f}_3&=&\frac{1}{\sqrt{n_\th}}\sin\half x_{n_\th} .\nonumber
\eea
The result, using identities similar to Eq.~(\ref{id}), is
\bea\label{hQM}
\tilde{H}&=&m_{V_0} n_V+m_N n_N +m_\th n_\th + g_0\sqrt{n_\th} \tan \half x_{n_\th}\ (n_N \bar{n}_V)\cr
&&-g_0\sqrt{n_\th+1} \tan\half x_{n_\th+1}\ (n_V\bar{n}_N).
\eea
In the $V/N\th$ sector this reduces to
\bea
\tilde{H}\equiv m_{V_0} n_V+m_N n_N +m_\th n_\th + g_0 \tan \half x\  (n_N n_\th -n_V),
\eea
which decouples $V$ from  $N$ and $\th$ and correctly reproduces the energy spectrum of Eq.~(\ref{evs}).
Note that because of the definition of $x$ in Eq.~(\ref{x}), $\tilde{H}$ is a function of $g_0^2$ rather than
$g_0$ itself.

\section{Field Theory in the $\boldsymbol{V/N\th}$ Sector}

In field theory the free and interaction Hamiltonians are, respectively,
\bea\label{H1QFT}
H_0&=&\int d\bp\left(m_{V_0} V_\bp^\dag V_\bp +m_N N_\bp^\dag N_\bp\right) +\int d\bk \omega_\bk a_\bk^\dag a_\bk,\cr
&&\\
H_1&=&\int d\bp d\bk h_\bk \left(V_\bp^\dag N_{\bp-\bk} a_\bk +a_\bk^\dag N_{\bp-\bk}^\dag V_\bp\right),\nonumber
\eea
where $\omega_\bk=(m_\th^2+\bk^2)^\half$ and\cite{Ghost}
\bea
h_\bk=\frac{g_0 \rho(\omega_\bk)}{(2\pi)^{3/2}(2\omega_\bk)^{1/2}},
\eea
where $\rho(\omega_\bk)$ is an appropriate cutoff.

As has been discussed in Ref.~\cite{Ghost}, it is again the case that above a certain critical value for
the renormalized coupling constant the unrenormalized coupling constant $g_0$ becomes imaginary. In field
theory the spectrum remains real, but a ghost state appears, with negative residue, and the Hamiltonian
becomes non-Hermitian (but ${\cal PT}$-symmetric).

Within the $V/N\th$ sector the $Q$-operator, satisfying (\ref{qH1}), then has the form\cite{Ghost}
\bea\label{QQFT}
Q=\int d\bp d\bk \gamma_\bk \left(V_\bp^\dag N_{\bp-\bk} a_\bk -a_\bk^\dag N_{\bp-\bk}^\dag V_\bp\right),
\eea
where $\gamma_\bk$, like $h_\bk$, is imaginary.
Again within the $V/N\th$ sector, $Q^2$ turns out to be
\bea
Q^2=\beta^2\int d\bp V_\bp^\dag V_\bp - \int d\bp d\bk_1 d\bk_2 \gamma_{\bk_1}\gamma_{\bk_2}
a_{\bk_1}^\dag N_{\bp-\bk_1} N_{\bp-\bk_2} a_{\bk_2},
\eea
where $\beta^2\equiv - \int d\bk\ \gamma_\bk^2$.

These are the only two independent structures, since it is easily seen that $Q^3=\beta^2 Q$
within the $V/N\th$ sector, i.e. ignoring terms that vanish when acting on a state in this sector.
Thus the series for $e^Q$ can be readily summed, to give
\bea
e^Q=1+c_1 Q+ c_2 Q^2.
\eea
where
\bea
c_1&=&\frac{\sinh\beta}{\beta},\nonumber\\
c_2&=&\frac{\cosh\beta-1}{\beta^2},
\eea
satisfying the important identity
\bea
c_1^2=2c_2+c_2^2\beta^2.
\eea
Inserting the above expression for $e^Q$ into Eq.~(\ref{qH1}) yields the condition
\bea\label{C}
c_1 \mu_\bk \gamma_\bk = c_2\beta_2 \gamma_\bk + \frac{c_1^2}{c_2} h_\bk ,
\eea
where $\mu_\bk= \omega_\bk+m_N-m_{V_0}$ and $\beta_2\equiv -\int d_\bk \gamma_\bk h_\bk$.
The other condition, that $\gamma_\bk$ be imaginary, has already been assumed. In principle $\gamma_\bk$ is found from this equation together with the defining relations
for $\beta^2$ and $\beta_2$.

We have again verified these calculations using the condition (\ref{qH}) rather than
(\ref{qH1}). This amounts to verifying the relation
\bea
2H_1&=&c_1[Q,H]+c_1^2QHQ -c_2\{Q^2,H\}\nonumber\\ &&-c_1c_2(Q^2HQ-QHQ^2)-c_2^2 Q^2H Q^2
\eea
and equating coefficients of the various operators that arise.

There are in fact just three of these, namely
\bea
{\cal O}_1&=&V_\bp^\dag V_\bp,\nonumber\\
{\cal O}_2&=& a_{\bk_1}^\dag N_{\bp-\bk_1} N_{\bp-\bk_2}^\dag a_{\bk_2},\\
{\cal O}_3&=& V_\bp^\dag N_{\bp-\bk} a_\bk +a_\bk^\dag N_{\bp-\bk}^\dag V_\bp.\nonumber
\eea
Thus in the expression on the RHS we require that the coefficients of ${\cal O}_1$ and ${\cal O}_2$
vanish, and that that of ${\cal O}_3$ match the coefficient in 2$H_1$. An important
ingredient in the manipulations is an identity obtained by integrating Eq.~(\ref{C}) with respect to
$\int d\bk \gamma_\bk$, namely
\bea\label{D}
c_1\int d\bk \gamma_k^2 \mu_\bk = -\beta_2\left( c_2 \beta^2+\frac{c_1^2}{c_2}\right).
\eea
To construct the equivalent Hermitian Hamiltonian $\tilde{H}$ we need $e^{\half Q}$, which in this case reads
\bea
e^{\half Q}=1+\tilde{c}_1 Q+ \tilde{c}_2 Q^2,
\eea
where
\bea
\tilde{c}_1&=&\frac{\sinh\half\beta}{\beta},\nonumber\\
\tilde{c}_2&=&\frac{\cosh\half\beta-1}{\beta^2},
\eea
satisfying
\bea
\tilde{c}_1c_1-\beta^2 \tilde{c}_2 c_2=c_2.
\eea
Again only the three structures ${\cal O}_1$, ${\cal O}_2$ and ${\cal O}_3$ are produced,
but this time the coefficient of ${\cal O}_3$ vanishes, leaving
\bea\label{hQFT}
\tilde{H}&=&H_0+\frac{\beta_2}{\beta}\tanh\half\beta\ \int d\bp V_\bp^\dag V_\bp\nonumber\\
&&+2\frac{\cosh\half\beta-1}{\beta\sinh\beta}\int d\bp d\bk_1 d\bk_2 a_{\bk_1}^\dag N_{\bp-\bk_1} N_{\bp-\bk_2}^\dag a_{\bk_2}\times\\
&&\hspace{2cm}\times\ [
\cosh\half\beta(\gamma_{\bk_1}h_{\bk_2}+\gamma_{\bk_2}h_{\bk_2})-\beta_2\tilde{c}_2 \gamma_{\bk_1}\gamma_{\bk_2}]\nonumber
\eea
We will make a more detailed comparison with the corresponding quantum-mechanical result,
Eq.~(\ref{hQM}), in the next section, but the general features are clear: the $V$ and $N\th$ sectors
are again separated, with a term that renormalizes the mass of the $V$ particle together with a four-point
interaction between $N$ and $\th$.
\section{Discussion}

We can find the quantum-mechanical limit of Eq.~(\ref{hQFT}) by ignoring all momentum subscripts and integrals
and identifying $\gamma=x$, by comparison of the two expressions for $Q$, Eqs.~(\ref{QQM}) and (\ref{QQFT}).
By comparison of the two expressions for $H_1$, Eqs.~(\ref{H1QM}) and Eqs.~(\ref{H1QFT}), we further
identify $h=g_0$.
Recall that the quantum-mechanical result is only valid for $g_0$ real. Thus we have to continue the field-theory
expressions to this regime, where $\gamma$ is real.
Thus $\beta^2=-x^2$, so that $\beta=\pm ix$, while $\beta_2=-g_0 x$. It follows almost immediately
that the coefficient of $V^\dag V$ reduces correctly to $-g_0 \tan\half x$, while the coefficient,
$+g_0 \tan\half x$, of $n_N n_\th$ follows after a little algebra.

Perhaps the most important feature of our work is that we can use Eqs.~(\ref{qH1}) or (\ref{grev}) for
both real and imaginary $g_0$.
In the case when $g_0$ is imaginary and $H$ is non-Hermitian, this is equivalent to the usual equation for
quasi-Hermiticity, Eq.~(\ref{qH}). Then $Q$ is Hermitian so that the relation (\ref{h}) represents a similarity transformation rather
than a unitary transformation, and consequently the metric of Hilbert space is changed. However, in the case
when $g_0$ is real, and the original $H$ is actually Hermitian, we can still derive a $Q$ operator from Eqs.~(\ref{qH1}) or (\ref{grev}).
In this case $Q$ is anti-Hermitian and the relation (\ref{h}) is a unitary transformation relating one Hermitian
Hamiltonian to another, with no change of the standard Hilbert space metric\footnote{The metric operator $\eta$ is
only equal to $e^{-Q}$ for $Q$ Hermitian. The general relation is $\eta=e^{-\half Q^\dag}e^{-\half Q}$, which instead reduces to the
identity for $Q$ anti-Hermitian.}. The common feature of both cases is that,
whereas the initial Hamiltonian is linear in $g_0$, the equivalent Hamiltonian is a function of $g_0^2$.
Thus, using a method inspired by the study of pseudo-Hermitian Hamiltonians, we have produced a Hermitian four-point interaction
Hamiltonian equivalent to the original Hermitian trilinear interaction Hamiltonian. This equivalent Hamiltonian $\tilde{H}$ is given
in Eq.~(\ref{hQFT}), where $\beta$ is real in the ghost regime, but pure imaginary in the Hermitian regime of $H$.

In graphical terms it is not obvious how $\tilde{H}$ is constructed. To order $g_0^2$ it amounts to contracting the $V$ propagator
in the $N\th$ scattering amplitude to a point, but it also takes into account higher-order loop diagrams in a complicated way.


\end{document}